\documentclass[twocolumn,showpacs,preprintnumbers,psfig,pre,superscriptaddress]{revtex4}

\usepackage{graphics}
\usepackage{epsfig}

\begin{document}

\title{Solvent-induced micelle formation in a hydrophobic interaction model}
\author{S. Moelbert}
\affiliation{Institut de th\'eorie des ph\'enom\`enes physiques, Ecole 
polytechnique f\'ed\'erale de Lausanne, CH-1015 Lausanne, Switzerland}

\author{B. Normand}
\affiliation{D\'epartement de Physique, Universit\'e de Fribourg, CH-1700 
Fribourg, Switzerland} 

\author{P. De Los Rios}
\affiliation{Institut de th\'eorie des ph\'enom\`enes physiques, Ecole 
polytechnique f\'ed\'erale de Lausanne, CH-1015 Lausanne, Switzerland}

\date{\today}

\begin{abstract}
We investigate the aggregation of amphiphilic molecules by adapting the 
two-state Muller-Lee-Graziano model for water, in which a solvent-induced 
hydrophobic interaction is included implicitly. We study the formation of 
various types of micelle as a function of the distribution of hydrophobic 
regions at the molecular surface. Successive substitution of non-polar 
surfaces by polar ones demonstrates the influence of hydrophobicity on 
the upper and lower critical solution temperatures. Aggregates of lipid 
molecules, described by a refinement of the model in which a hydrophobic 
tail of variable length interacts with different numbers of water molecules, 
are stabilized as the length of the tail increases. We demonstrate that the 
essential features of micelle formation are primarily solvent-induced, and 
are explained within a model which focuses only on the alteration of water 
structure in the vicinity of the hydrophobic surface regions of 
amphiphiles in solution.

\end{abstract}

\maketitle

\section{Introduction}
\begin{figure}[b!]
\includegraphics[width=7cm]{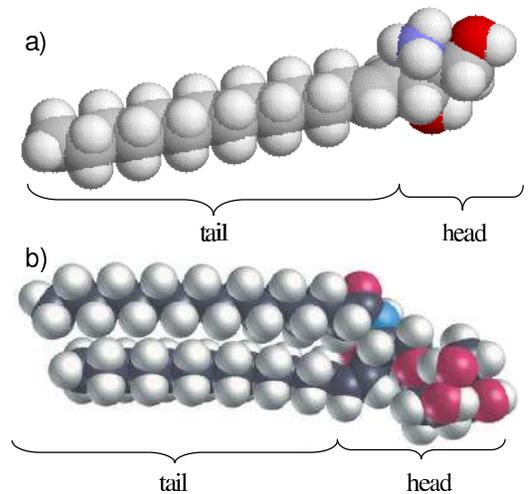}
\caption{Space-filling models of typical lipid molecules: (a) a sphingolipid 
(sphingosine, $C_{18}H_{37}O_2N$), with one non-polar hydrocarbon tail; (b) 
a glycolipid ($C_{40}H_{72}O_{8}N$), with two tails. The relatively large, 
polar head-groups contain in addition oxygen and nitrogen atoms.   }
\label{FigGlycoLipid}
\end{figure}

In an aqueous solution, the hydrophobic parts of amphiphilic molecules 
tend to separate themselves from water molecules by forming aggregates, 
such as micelles and microemulsion droplets. The simplest amphiphilic 
structure occurs if the polar and hydrophobic parts of the amphiphilic 
molecule are well separated into head and tail regions. Motivated by 
the structure of sphingolipids and glycolipids (Fig.~\ref{FigGlycoLipid}), 
we will refer to this type of molecule as `lipids' (although we note here 
that this generic structure is not common to all classes of lipid). For 
molecules such as those in Fig.~\ref{FigGlycoLipid}, micelles consist of 
a polar outer surface and a hydrophobic core which contains all the tails. 
By contrast, amphiphilic species whose polar and hydrophobic regions are 
distributed over the entire molecule, rather than being clearly separated, 
aggregate to form more general assemblies for the purpose of minimizing 
the hydrophobic area per molecule exposed to the aqueous phase. We  will 
refer to this general category of mixed HP molecules as `amphiphiles.' 
Poly(N-isopropylacrylamide), a polymer belonging to this class, exhibits 
a phase transition at a lower critical solution temperature (LCST) from a 
homogeneous solution, where the polymers are completely soluble, to a 
system of two separated phases~\cite{azevedo}. At an upper critical 
solution temperature (UCST) the organic phase disaggregates, and above 
this temperature the amphiphilic molecules are again soluble due to 
entropic effects~\cite{rebelo}. Substitution of polar by hydrophobic 
monomers in amphiphiles of given length leads to alterations of the 
critical solution temperatures which depend on the size of hydrophobic 
surface regions~\cite{rebelo}.

Micelle formation, and the aggregation of amphiphilic molecules in general, 
may be treated as a phase separation occurring at a critical micellar 
concentration (CMC) which describes the density of amphiphiles where the 
system enters the two-phase region~\cite{stillinger-benNaim, leibler}. 
Above the CMC, amphiphilic molecules in aqueous solutions self-associate, 
forming small aggregates to decrease the net contact between their 
hydrophobic surfaces and the surrounding solvent. A small fraction of the 
lipid molecules remains free in the solution, with a concentration close 
to the CMC value. As the concentration increases, the onset of a 
semi-dilute regime is found, where the system may be considered as a 
solution of relatively few water molecules dissolved in an amphiphilic 
medium~\cite{israelachvili}.

For lipid molecules, the CMC decreases as the length of the hydrophobic 
chain increases~\cite{tanford,chen,lang,wenzel}, indicating a stabilization 
of aggregates as a consequence of the stronger net repulsion between the 
tail and the surrounding water. For the same reason, the LCST is thought 
to decrease as the tail grows, as suggested in Ref.~\cite{rebelo}. One 
means of probing the nature of the phase diagram and the effective 
hydrophobic interactions would be by systematic alteration of the 
polarity of the amphiphilic polymers in solution. Many theoretical 
and experimental studies have been conducted to describe the various 
aggregation phases of amphiphilic molecules in aqueous 
solutions~\cite{israelachvili,wennerstrom,tanford,chen,lang,wenzel,corti,
goldstein,stillinger-benNaim,larson,leibler,bhatta1,bhatta2,bhatta3,tiddy,
gompper}. However, no comprehensive investigations have yet been performed 
concerning the influence of the distribution of polar groups in amphiphiles 
on the aggregation phase diagram, or concerning the mechanism underlying 
the process of self-aggregation. 

The aim of this study is to substantiate experimental results indicating 
a decrease in LCST as the degree of hydrophobicity increases, and to 
analyze the dependence on density and hydrophobicity of the aggregation 
phase diagram. We will conclude that the principal properties of 
amphiphiles in aqueous solutions are solvent-induced, in that they are 
explained by alterations in the structure of liquid water in the vicinity 
of the hydrophobic regions of solute particles. We begin by introducing a 
simple hydrophobic-polar (HP) description (Sec.~\ref{secHPModel}) of 
amphiphilic solute particles (Sec.~\ref{secHPModelMic}) on a cubic lattice, 
and then extend the model to describe lipid particles of varying tail 
length (Sec.~\ref{secLipid}). In Sec.~\ref{secResultsMic} we investigate 
the changes in the phase diagram associated with an increasing proportion 
of polar groups in amphiphilic molecules and with changes in their 
distribution. Sec.~\ref{secResultsLipid} presents a similar analysis for 
the formation of lipid micelles in aqueous solutions. In 
Sec.~\ref{secDiscussion} we discuss the implications of our results 
and provide a brief conclusion.

\section{Model and Methods}
\label{secModel}

\subsection{Water Structure}  
\label{secHPModel}

\begin{figure}[t!]
\includegraphics[width=6.5cm]{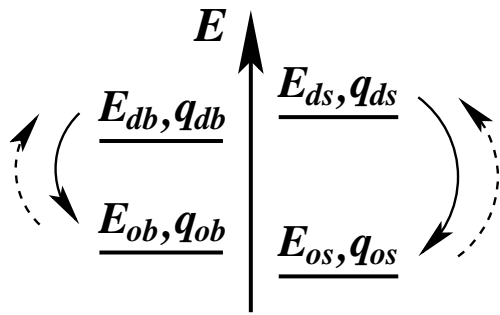}
\caption{Energy levels of a water site in the bimodal MLG model. The states 
are denoted $os$ = ordered shell (cage conformation),  $ds$ = disordered 
shell, $ob$ = ordered bulk, and $db$ = disordered bulk.}
\label{FigMLG}
\end{figure}

The solvation of amphiphiles in aqueous solutions and their self-association 
into micellar aggregates are genreally considered as a consequence of the 
effective hydrophobic interaction between polar water and the non-polar 
regions of the solute molecules~\cite{kauzmann}. The unique properties of 
the aqueous medium which generate this interaction arise from the ability 
of water molecules to form strong hydrogen bonds both among themselves and 
with the polar groups of solute molecules. The formation and disruption of 
extended hydrogen-bonded networks leads directly to the delicate balance of 
enthalpic and entropic contributions to the solvent free energy which is 
responsible for the existence of a closed-loop aggregation regime in many 
solutions of polar molecules. This behavior is encapsulated in an adapted 
version of the model of Muller, Lee and Graziano (MLG)~\cite{muller,lee}, 
whose key features are as follows: consideration of a coarse-grained 
system whose sites contain clusters of water molecules of a size which 
matches that of the solute particles; a bimodal distribution of water 
clusters reflecting ``ordered'' sites with mostly intact hydrogen bonds, 
and ``disordered'' sites with relatively fewer intact hydrogen bonds; a 
further subdivision into ``shell'' sites neighboring the solute particles 
and ``bulk'' sites which are not disrupted by proximity to a polar 
surface; a set of microscopic energy and degeneracy parameters for these 
sites whose sequences are determined by the fact that breaking of hydrogen 
bonds increases site enthalpy but simultaneously raises the site degeneracy 
(number of orientational degrees of freedom of the water molecules), as 
represented in Fig.~\ref{FigMLG}. Full details of the foundations and 
qualitative properties of the model can be found in Refs.~\cite{moelbert1} 
and \cite{moelbert2}.

The microscopic origin of the energy and degeneracy parameters for the four 
different types of water site (ordered/disordered and shell/bulk) in the 
adapted MLG model, based on both experimental~\cite{privalov,deJong,pertsemil} 
and theoretical~\cite{stillinger80,ben-naim2,nemethy,ludwig} analysis, is 
discussed in detail in Refs.~\cite{silverstein2,moelbert1}. Here we 
highlight the competition of enthalpy and entropy terms using only the 
example of the ordered states. Although the insertion of a hydrophobic 
molecule leads to a destruction of local hydrogen-bonding, at low 
temperatures the water molecules are found to rearrange in a cage-like 
structure around the solute molecule, which because of the orientational 
effect of the polar surface is formed by stronger hydrogen bonds. The 
consequent net energy reduction results in dissolution of the solute 
particle~\cite{privalov,deJong,pertsemil}. At higher temperatures, 
however, the additional entropic contributions available from the 
bulk solvent favor a minimization of local water restructuring, which 
drives the aggregation of hydrophobic solute particles to minimize their 
total surface exposed to water. 

While it is possible to find perfect cages of hydrogen bonds for small 
hydrophobic particles, the formation of complete cages around large solute 
particles is prevented sterically. However, such large solute particles, 
and also their aggregates, may instead be surrounded by a number of partial 
cages, depending on their shape and surface roughness~\cite{nemethy}. 
Atomic groups exposed at the surfaces of large molecules and aggregates 
present locally curved surface structures \cite{kita} which allow for 
a confined formation of partial cages. The lipid molecules we analyze 
(see Fig.~1) have a characteristic lengthscale of order 0.3 nm for the 
diameter of the chain and any side groups, 0.5 nm for the polar head, 
and a maximum length of 2-3 nm for the longer chains. The chain and head 
diameters are the lengthscales which determine the curvature of the 
molecule, and thus the possibility for cage formation. The considerations 
of the previous paragraph, specifically low-temperature solution, are 
then expected to remain valid for amphiphilic and lipid molecules, 
and this is confirmed by the success of models based on these concepts in 
describing large molecular species~\cite{delos,caldarelli,delos2}. That 
the analysis is applicable for solute particles with a certain degree of 
curvature and surface roughness does not appear to be a significant 
restriction for molecules in the size range of most interest for 
amphiphilic and lipid characteristics.

Returning to the parameters of Fig.~\ref{FigMLG}, we stress that the 
properties of the sysytem are critically dependent on the sequences $E_{ds} 
> E_{db} > E_{ob} > E_{os}$ of energy levels and $q_{ds} > q_{db} > q_{ob} 
> q_{os}$ of the corresponding degeneracies, which describe the entropy. 
These sequences are confirmed by a range of experimental measurements, and 
indeed if they are not maintained the solution does not exhibit a closed-loop 
aggregation regime. However, the results of the calculations to follow are 
not particularly sensitive to the exact values of the differences between 
these parameters, and are of course independent of their absolute values. 
We have used the energy values $E_{ds} = 1.8,\ E_{db} = 1.0,\ E_{ob} = 
-1.0$, and $E_{os} = -2.0$, which are thought to be qualitatively 
representative for aqueous solutions, and which have been successful in 
describing different types of solution~\cite{delos2,silverstein2,
moelbert1,moelbert2}. The respective degeneracies, normalized to a 
non-degenerate ordered shell conformation, are taken to be $q_{ds} = 49,\ 
q_{db} = 40,\ q_{ob} = 10$ and $q_{os} = 1$~\cite{silverstein2}. These 
relative values have been found to be appropriate for reproducing the 
phenomenology of hydrophobic interactions~\cite{silverstein2,moelbert1}, 
protein denaturation~\cite{delos,caldarelli}, swelling of 
biopolymers~\cite{delos2}, and cosolvent effects on solubility of 
hydrophobic particles~\cite{moelbert2,moelbert3}. Precise values of 
the microscopic parameters may in fact be refined by comparison with 
experimental measurements to yield semi-quantitative agreement for 
different solutions~\cite{delos2,silverstein2,moelbert1,moelbert3}. 
The energy scale is correlated directly with a relative temperature 
scale, which we define as $k_BT\equiv \beta^{-1}$.

\subsection{Amphiphiles}  
\label{secHPModelMic}
To include not only purely hydrophobic solute particles but also amphiphilic 
molecules with varying conformations of polar and non-polar regions, as in 
the experiments of Ref.~\cite{azevedo}, we represent the particles as cubes 
on which each face may be either polar or hydrophobic. A neighboring water 
site, which is homogeneously polar, interacts only with the side of the 
particle which is oriented in its direction. If this side represents a 
polar group, the neighboring water site is in a bulk state, whereas if 
the side is hydrophobic the water site is considered a shell state. Thus 
polar sides and water are considered as having the same effect on neighboring 
water molecules. However, polar faces do not contribute to the free energy. 
In this coarse-grained model, one side of a site may represent more than 
one chemical group, and thus corresponds to a net characterization of the 
surface area of the solute molecule under consideration.  \\

Hydrophobic solute particles are generally larger than water, and thus a 
water site in the model consists of a group of molecules. On a lattice 
where each site has $z$ nearest neighbors, the energy of a system of $N$ 
sites, occupied either by amphiphiles ($n_i=0$) or by water ($n_i=1$), is 
given by the Potts-like Hamiltonian
\begin{eqnarray}\nonumber
H[&\!\!\!\{n_i\},&\!\!\!\{\sigma_i\},\{k\}]=\\ \nonumber
&&\sum_{i=1}^{N} \frac{n_i}{z} \!\left[\! (E_{os} \tilde\delta_{i, 
\sigma_{os}}\!+\!E_{ds}\tilde\delta_{i, \sigma_{ds}})\!\sum_{\langle 
ji \rangle} (1\!-\!\lambda_{i,j(k)}) \right.\\
&&+\left.(E_{ob} \tilde\delta_{i, \sigma_{ob}}\!+ E_{db} \tilde\delta_{i, 
\sigma_{db}}) \sum_{\langle ji \rangle} \lambda_{i,j(k)} \right],
\label{Ham}
\end{eqnarray}
where $\lambda_{i,j(k)}$ is a side variable depending on each of the 
nearest neighbors $j$ of site $i$, and takes the value $1$ if the 
neighboring side of site $j$ is a polar side of a solute particle or water, 
and $0$ otherwise. The variable $k$ takes values from $1$ to $n_{c,j}$, 
where $n_{c,j}$ enumerates all equivalent orientations of site $j$. On 
a cubic lattice, the total number of sides of a solute particle is $z = 
6 = N_P + N_H$, where $N_H$ is the number of hydrophobic cube faces.
Thus for the cubic lattice $n_{c,j} = 1$ for water and for 
solute particles with $N_H = 6$, $n_{c,j} = 6$ for $N_H = 5$, $n_{c,j} = 
3$ for $N_H = 4$ ($N_H = 2$) if the two polar (hydrophobic) sides are 
opposite to each other, $n_{c,j} = 12$ for $N_H = 4$ ($N_H = 2$) if they 
are adjacent, $n_{c,j} = 8$ for $N_H = 3$ if the three polar sides are 
are all adjacent to each other, and $n_{c,j} = 12$ for $N_H = 3$ if two 
of them are opposite to each other. 

Because a water side may be in one of $q$ different states, 
$\tilde\delta_{i, \sigma_{os}}$ is $1$ if it is in one of the $q_{os}$ 
ordered shell states and $0$ otherwise, and $\tilde\delta_{i, \sigma_{ds}}$ 
is $1$ if it is in one of the $q_{ds}$ disordered shell states and $0$ 
otherwise. Analogous considerations apply for the bulk states. 

We note that for completely hydrophobic solute particles ($N_P = 0$) 
Eq.~(\ref{Ham}) is not equivalent to the Hamiltonian in 
Ref.~\cite{moelbert1}, because the sites are treated differently. In 
Eq.~(\ref{Ham}) each face of a water site $i$ contributes to the free 
energy, and a sum is performed over all pairs of faces in contact with 
each other, as opposed to over individual sites. The representation in 
terms of cube sides is adopted here for consistent description of lipid 
molecules in Sec.~\ref{secLipid}. However, we will find in 
Sec.~\ref{secResultsMic} that the qualitative differences between 
the two Hamiltonians for fully hydrophobic solute particles are small. 

To determine the canonical partition function, a sum is performed over 
the state configurations $\{\sigma_i\}$. By taking into account the 
possible orientations of the amphiphilic particles through the variable 
$k$, the canonical partition function of the system of $N$ sites may be 
expressed as
\begin{equation} 
Z_N=\sum_{\{n_i\}} 
\prod_{i=1}^{N} \prod_{k=1}^{n_{c,j}}
Z_b^{n_i \frac{1}{z}\sum_{\langle ji \rangle} \lambda_{i,j(k)}} 
Z_s^{n_i \frac{1}{z}\sum_{\langle ji \rangle} (1 - \lambda_{i,j(k)})},
\label{partMicHP}
\end{equation}
where $Z_{\sigma} = q_{o\sigma}e^{-\beta E_{o\sigma}}+q_{d\sigma} e^{-\beta 
E_{d\sigma}}$ for the shell ($\sigma \equiv s$) and bulk ($\sigma \equiv b$) 
states of pure water sides. The grand canonical partition function of the 
system for variable solute particle number is then
\begin{equation} 
\Xi = \sum_{N} e^{\beta \mu N_w } Z_N = \sum_{\{n_i\}} e^{-\beta H_{\rm 
eff}^{\rm gc}[\{n_i\}]} ,
\label{eq:gcZMic}
\end{equation} 
where $\mu$ represents the chemical potential associated with the insertion 
of water and $N_w$ denotes the number of water sites. Although the explicit 
terms of the model describe solely the states of water molecules in solution, 
it contains implicitly all multi-particle interactions between hydrophobic 
solute molecules~\cite{moelbert1}.

The coexistence regions are characterized by measuring the UCST and the 
LCST for various numbers $N_P$ of polar sides per particle. To investigate 
variations in the effective hydrophobic interactions due to changes in 
polarity of the solute particles, we  increase systematically the number 
of polar sides per particle and determine the coexistence region in each 
case.   

The extent of aggregate formation in the system is determined from the number 
density of contacts between two hydrophobic cube sides, $n_{H\!-\!H}$, 
between two polar sides, $n_{P\!-\!P}$, and between a polar and a 
hydrophobic side, $n_{P\!-\!H}$. The number of contacts in a randomly 
distributed system with the same particle density $\rho_p$ is also 
calculated for comparison. In a random solution of solute particles 
with $N_P$ polar faces, whose positions and orientations are completely 
independent, these probabilities are given in the thermodynamic limit by
\begin{eqnarray} \nonumber
p(H\!\!-\!\!H) &= &p(H)^2, \\ \nonumber
p(P\!\!-\!\!P) &= &p(P)^2, \\ 
p(P\!\!-\!\!H) &= &2\, p(H)\, p(P),
\end{eqnarray}
where $p(H)$ is the probability that a cube face is hydrophobic, $p(P)$ 
is the probability that it is polar, and by symmetry $p(H\!\!-\!\!P) = 
p(P\!\!-\!\!H)$. In a random system, the probabilities of occurrence of 
the different faces are independent of the neighboring sites, and are simply
\begin{eqnarray} \nonumber
p(H) &= &\frac{N_H}{z} \rho_p, \\
p(P) &= &(1-\rho_p)+\frac{N_P}{z} \rho_p.
\end{eqnarray}
If the contact densities $n_{H\!-\!H}$ and $n_{P\!-\!P}$, are larger than 
their probabilities of random occurrence, and $n_{P\!-\!H}$, is 
correspondingly smaller, the system has formed aggregates which reduce 
the number of hydrophobic sides exposed to water.

\subsection{Lipids}
\label{secLipid}
Micelles are generally formed by amphiphilic molecules, referred to here 
as lipids, which are composed of two distinct regions, the polar head and 
the hydrophobic tail. The length of the tail, which is typically composed 
of one or more hydrocarbon chains, is normally rather greater than the 
size of the polar head (Fig.~\ref{FigGlycoLipid}). Experimental 
observations~\cite{yalkowsky} suggest that lipids with tails shorter 
than $10$ carbon atoms are highly soluble in aqueous solutions, while 
those whose tails exceed approximately $20$ carbon atoms are almost 
completely insoluble. For those molecules with tail lengths in the 
intermediate range, which show the widest variety of surface-active 
properties, the tails are on average some three to five times longer 
than the dimensions of the polar head. (In lipids composed of two or 
more tails, this ratio is generally smaller.) As an example, the head 
size of the glycolipid in Fig.~\ref{FigGlycoLipid} is close to 
one third of the length of the 14-atom tail. We define an effective tail 
length $l$ as the ratio between the tail length and the head size of the 
lipid, so that a typical lipid is represented by $l$ values between two 
and five, which may also be fractional.  

For simple geometrical reasons, the total repulsion between such a tail 
and the water molecules surrounding its sides is significantly stronger 
than that for the small tail tip of the chain. We adapt the model described 
in Sec.~\ref{secHPModelMic} to include this aspect by assigning different 
energy levels to shell water clusters in contact with the sides of a solute 
particle compared to those in contact with the tip. Because the number of 
shell water molecules interacting with one side of a hydrophobic tail is 
approximately $l$ times that of those interacting with the tip, the energy 
associated with a site representing all of these water molecules is taken 
to be $l$ times that for a tip site. A cubic solute particle in the lipid 
model [represented in Fig.~\ref{FigMicPart3D}(a)] consists of a polar head 
(P) and a hydrophobic tail, which in turn is divided into the moderately 
hydrophobic tip (H), situated opposite the polar head, and the long, 
strongly hydrophobic sides of the tail (S). In the coarse-grained model, 
both the tip and each long side of the tail are represented by a face of 
the cubic particle, but the sides interact more strongly with a neighboring 
water site than does the tip.

The Hamiltonian of a system of $N$ sites on a cubic lattice, which are 
occupied either by water ($n_i=1$) or by a lipid molecule ($n_i=0$), is then 
\begin{eqnarray}
\nonumber
H[&&\!\!\!\!\!\!\{n_i\},\{\sigma_i\},\{k\}]=\\ \nonumber
&&\!\!\!\!\!\!\sum_{i=1}^{N} n_i[(E_{os,H} \tilde\delta_{i, \sigma_{os,H}}\! 
+\! E_{ds,H} \tilde\delta_{i, \sigma_{ds,H}})\frac{1}{z}\sum_{\langle ji 
\rangle} (1\!-\!\lambda_{i,j(k)}^2) \\ \nonumber
&&\!\!\!\!\!\!+(E_{os,S} \tilde\delta_{i,\sigma_{os,S}}\!+\!E_{ds,S} 
\tilde\delta_{i, \sigma_{ds,S}})\sum_{\langle ji \rangle}\!\frac{1}{2z}
\lambda_{i,j(k)} (1-\lambda_{i,j(k)})\\
&&\!\!\!\!\!\!+ (E_{ob} \tilde\delta_{i, \sigma_{ob}}\!+\!E_{db} 
\tilde\delta_{i,\sigma_{db}})\sum_{\langle ji \rangle}\frac{1}{2z}
\lambda_{i,j(k)}(1\!+\!\lambda_{i,j(k)})],
\label{Hamilt}
\end{eqnarray}
where again $\lambda_{i,j(k)}$ depends on the orientation state $k$ of the 
particle at each nearest-neighbor site $j$ of $i$. However, for the lipid 
model  $\lambda_{i,j(k)}$ takes the value $1$ if the neighboring face is 
polar ({\sl i.e.} water or the relevant side of a solute particle), $0$ 
if it represents a slightly hydrophobic tail tip H, and $-1$ if it 
represents a strongly hydrophobic tail side S. 
$k$ varies again from $1$ to $n_{c,j}$, where $n_{c,j} = 1$ for water and 
$n_{c,j} = 6$ for lipid particles. Because a polar side $i$ may be in one 
of $q$ different states, $\tilde\delta_{i, \sigma_{os,H}}$ is $1$ if site 
$i$ is occupied by a polar face in one of the $q_{os}$ ordered shell states 
of a H face and $0$ otherwise, and $\tilde\delta_{i, \sigma_{ds,H}}$ is $1$ 
if it is occupied by pure water in one of the $q_{ds}$ disordered shell 
states of a H face and $0$ otherwise. Analogous considerations apply for 
S faces and for the bulk states. We stress here that the Hamiltonians in 
Eqs.~(\ref{Ham}) and~(\ref{Hamilt}) provide a full description of the 
microscopic states of the amphiphilic and lipid systems. The macroscopic, 
thermodynamic behavior is contained implicitly in these microscopic 
states~\cite{moelbert1,moelbert2,moelbert3,delos,delos2,caldarelli,
muller,lee,silverstein1,silverstein2}, and the differences for such 
properties as the phase boundaries for varying chain length enter 
through the microscopic energy and degeneracy parameters.\\

As in Sec.~\ref{secHPModelMic} only water sites contribute to the free 
energy, and a polar face is considered having the same effect on a 
neighboring water site as water. We define 
\begin{eqnarray} \nonumber
{\cal S}_H &= &-\frac{1}{\beta} \ln \left[ q_{os,H} e^{-\beta E_{os,H}} 
+ q_{ds,H} e^{-\beta E_{ds,H}} \right], \\ \nonumber
{\cal S}_S &= &-\frac{1}{\beta} \ln \left[ q_{os,S} e^{-\beta E_{os,S}} 
+ q_{ds,S} e^{-\beta E_{ds,S}} \right], \\ 
{\cal B} &= &-\frac{1}{\beta} \ln \left[ q_{ob} e^{-\beta E_{ob}} 
+ q_{db} e^{-\beta E_{db}} \right].
\end{eqnarray}
For a system of $N$ sites, the canonical partition function may be expressed as
\begin{eqnarray} 
Z_N &\!\!\!= &\!\!\!\!\sum_{\{n_i\}}  \prod_{i=1}^{N} \prod_{k=1}^{n_{c,j}}
e^{-\beta  \frac{n_i}{2z}{\cal B}\sum_{\langle ji \rangle}\lambda_{i,j(k)}
(\lambda_{i,j(k)}+1)}          \\
&\!\!\!\times &\!\!\!\!e^{\!-\beta \frac{n_i}{z}[{\cal S}_H\! \sum_{\langle 
ji \rangle}\!\! (1\!-\!\lambda_{i,j(k)}^2) +   {\cal S}_S\! \sum_{\langle 
ji \rangle}\!\!\frac{1}{2}\lambda_{i,j(k)} (1\!-\!\lambda_{i,j(k)})  ]}\!\!, 
\nonumber
\end{eqnarray}
whence the grand canonical partition function for a system where the lipid 
density is not fixed is 
\begin{equation} 
\Xi = \sum_N e^{-\beta \mu N_w} Z_N = \sum_{\{n_i\}} e^{-\beta H_{\rm 
eff}^{\rm gc} [\{n_i\}]},
\end{equation}
with the effective, grand canonical Hamiltonian 
\begin{eqnarray} \nonumber
H_{\rm eff}^{\rm gc} [\{n_i\}] &= &\sum_{i=1}^{N} \sum_{k=1}^{n_{c,j}}[n_i(
{\cal S}_H \frac{1}{z}\sum_{\langle ji \rangle} (1-\lambda_{i,j(k)}^2)
-\mu)\\ \nonumber
&&+ {\cal B} n_i\sum_{\langle ji \rangle}\frac{1}{2z}\lambda_{i,j(k)}
(1\!+\!\lambda_{i,j(k)}) \\
&&+ {\cal S}_S n_i\sum_{\langle ji \rangle}\frac{1}{2z}\lambda_{i,j(k)} 
(1-\lambda_{i,j(k)})].  
\label{eq:HeffMic}
\end{eqnarray}

To measure the formation of aggregates in the lipid system, the number of 
contacts between different faces is again determined and compared with the 
number of contacts in a randomly distributed system of the same particle 
density $\rho_p$. In a random solution of solute particles, whose positions 
and orientations are completely independent, these probabilities are given 
in the thermodynamic limit by 
\begin{eqnarray} \nonumber
p(\alpha\!\!-\!\!\alpha) & = & p(\alpha)^2, \\ \nonumber
p(\alpha\!\!-\!\!\beta) & = & 2\, p(\alpha)\, p(\beta) \:\, = \:\, 
p(\beta\!\!-\!\!\alpha),  
\end{eqnarray}
where $\alpha, \beta \equiv H, P, S$. Here $p(H)$ is the probability that 
the adjacent side of the nearest neighbor is slightly hydrophobic, $p(P)$ 
the probability that it is polar, and $p(S)$ the probability that it 
is strongly hydrophobic. The probabilities are given in general by
\begin{eqnarray} \nonumber
p(H) &= &\frac{N_H}{z} \rho_p, \\ \nonumber
p(P) &= &(1-\rho_p)+\frac{N_P}{z} \rho_p, \\
p(S) &= &\frac{N_S}{z} \rho_p,
\end{eqnarray}
although henceforth we will consider only the values $N_H = N_P = 1$, 
$N_S = 4$. Aggregate formation has occurred in this system if $n_{H\!-\!H}$, 
$n_{P\!-\!P}$, and $n_{S\!-\!S}$ are larger than their random expectation 
values, and $n_{P\!-\!H}$ and $n_{P\!-\!S}$ are correspondingly smaller.  

The absolute energy levels and the degeneracies of water sites facing S 
sides are higher because this site represents the number of water molecules 
contained in a shell site of H multiplied by the factor $l$. The 
effective energy of a S shell site is thus obtained from that for a H 
shell site, for both ordered and disordered states, using
\begin{equation}
E_{os, S} = l \, E_{os, H}, \;\;\;\; E_{ds, S} \; = \; l \, E_{ds, H},
\end{equation}
where $E_{os, H}$ ($E_{ds, H}$) is the energy of an ordered (disordered) 
water site in the shell of the tip (H) and $E_{os, S}$ ($E_{ds, S}$) that 
of an ordered (disordered) shell water site of the tail side (S). Under 
the assumption that neighboring water sites are rather independent, the 
total number of configurations for two sites may be approximated by the 
product of their numbers of configurations. Thus, the number of 
configurations of a S shell site is related to that of a corresponding 
H shell site by
\begin{equation}
q_{os, S} = q_{os, H}^l, \;\;\;\; q_{ds, S} \; = \; q_{ds, H}^l.
\end{equation}
The parameter values of the energy levels, and the degeneracies of bulk 
water and H shell sites, are chosen as described in Sec.~\ref{secHPModel}.

\subsection{Methods} 
\label{secMethods}

Our interest is focused on the orientation and location of amphiphilic 
molecules in solution, which may be captured by molecular-level simulations. 
We thus restrict our considerations to Monte Carlo studies, because the 
processes involved depend strongly on local, spatial effects which are 
neglected in mean-field calculations. As in Ref.~\cite{moelbert1}, we use 
Monte Carlo simulations to detect the aggregation of amphiphilic solute 
particles as a function of temperature and solute concentration. We work 
with a system of $30\times30\times30$ sites on a cubic lattice with random 
initial particle distributions and periodic boundary conditions, using a 
Metropolis algorithm for sampling of the configuration space. The numbers 
of relaxation ($100\,000$) and measurement ($1\,000\,000$) steps are similar 
to those in our previous studies~\cite{moelbert1}. The closed-loop 
coexistence curves in the $\rho_p$-$T$ phase diagram are obtained from 
the transitions determined by increasing the temperature at fixed chemical 
potential (grand canonical sampling), which results in a sudden density 
jump at the transition temperature, and from the corresponding solute 
particle densities.

Analysis of the properties of a lipid-water system at the molecular 
level is possible by similar Monte Carlo simulations using the model of 
Sec.~\ref{secLipid}. However, the procedure described above must be 
redefined in one respect. In the Metropolis algorithm, the relative 
transition probability to a configuration $\{n'_i\}$ from a previous 
one $\{n_i\}$ depends on the difference in free energy of the two 
configurations according to
\begin{equation}
r = e^{-\beta (H_{\rm eff}^{\rm gc}[\{n'_i\}] - H_{\rm eff}^{\rm 
gc}[\{n_i\}])}.
\end{equation}
The free-energy difference must be calculated for two states with the 
same number of molecules. One step of the simulation procedure consists 
either in rotation of a solute particle or in a site exchange between 
two randomly chosen sites. The only contributions to the difference in 
free energy are then the energy change of the sites concerned, and of 
their nearest-neighbor sides. If a side changes from P to S or vice versa, 
the bulk and shell states of neighboring water sites must contain the same 
number of water molecules to be comparable. In this case, we attribute the 
bulk energies $E_{ob, S} = l \, E_{ob}$, $E_{db, S} = l \, E_{db}$, and the 
respective degeneracies $q_{ob, S} = q_{ob}^l$, $q_{db, S} = q_{db}^l$, to 
the relevant water sites when calculating the probability $r$. The index S 
refers to the comparison of this bulk site with a S shell site.

\section{Results}

\subsection{Amphiphiles}
\label{secResultsMic}

\begin{figure}[t!]
\includegraphics[width=8.5cm]{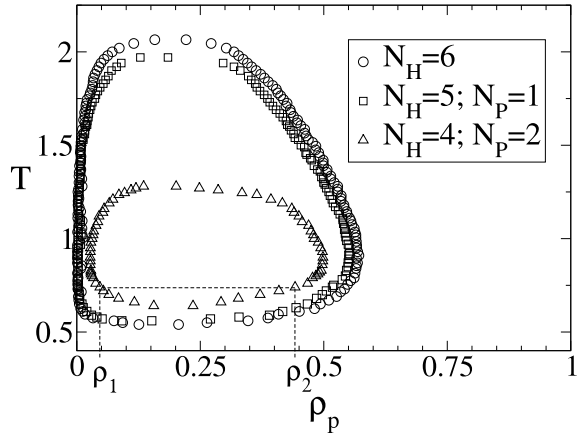}
\caption{$\rho_p$-$T$ phase diagram for micelle formation in the 3D, HP 
model for different numbers of polar sides per solute particle. The 
coexistence region is reduced, and aggregation suppressed, as the number 
of hydrophobic sides substituted by polar sides increases. At a given 
dimensionless temperature $T_{LCST}<T<T_{UCST}$ the system is homogeneous 
for solute particle concentrations below the CMC ($\rho_1$), while above 
this value it separates into two phases of densities $\rho_1$ and $\rho_2$. }
\label{FigBouleMicHP}
\end{figure}

We begin by attempting to capture the qualitative behavior 
of amphiphilic molecules with varying conformations of polar and non-polar 
regions. Fig.~\ref{FigBouleMicHP} shows the $\rho_p$-$T$ phase diagram 
obtained from the simple model of Sec.~\ref{secHPModelMic} for different 
numbers of polar sides per particle. The densities represent volume 
fractions, and are therefore dimensionless. In the coexistence regime 
the amphiphilic particles aggregate, minimizing the contact of their 
hydrophobic regions with water and with polar solute segments. Outside 
this region the amphiphiles are soluble at all densities below the LCST 
and above the UCST. For temperatures between these values, the amphiphiles 
are soluble only at very low densities, while at the CMC, $\rho_1$, the 
particles aggregate and the system separates into two phases: nearly pure 
water (of particle density $\rho_1$) and an amphiphilic phase of density 
$\rho_2$. With increasing polarity, the solubility of the solute particles 
is enhanced and the CMC increases.

As expected, the coexistence region is reduced as the number of polar sides 
increases~\cite{rebelo,lang}. If the system represents a solution of purely 
hydrophobic particles ($N_P = 0$ and $N_H = 6$), the coexistence curve is 
almost identical to that found in the HP model in Ref.~\cite{moelbert1} 
(see Sec.~\ref{secHPModelMic}). Substitution of one hydrophobic side per 
particle by a polar side ($N_P = 1$ and $N_H = 5$) leads to a decrease 
of the UCST and a slight increase of the LCST, and to a small overall 
suppression of the temperature and density range of the coexistence 
region. In this case, which may be taken to represent simplified lipid 
molecules (discussed in Sec.~\ref{secDiscussion}), the effect is rather 
moderate. However, substitution of a second hydrophobic side per particle 
($N_P = 2$ and $N_H = 4$) reduces the coexistence region dramatically 
(Fig.~\ref{FigBouleMicHP}). When substituting three or more hydrophobic 
sides by polar ones, the solute particles become soluble at all 
temperatures and no aggregation phase transition is observed.  

\begin{figure}[!t]
\begin{minipage}[t]{.46 \linewidth}
\center\epsfxsize=5.5cm \epsfbox{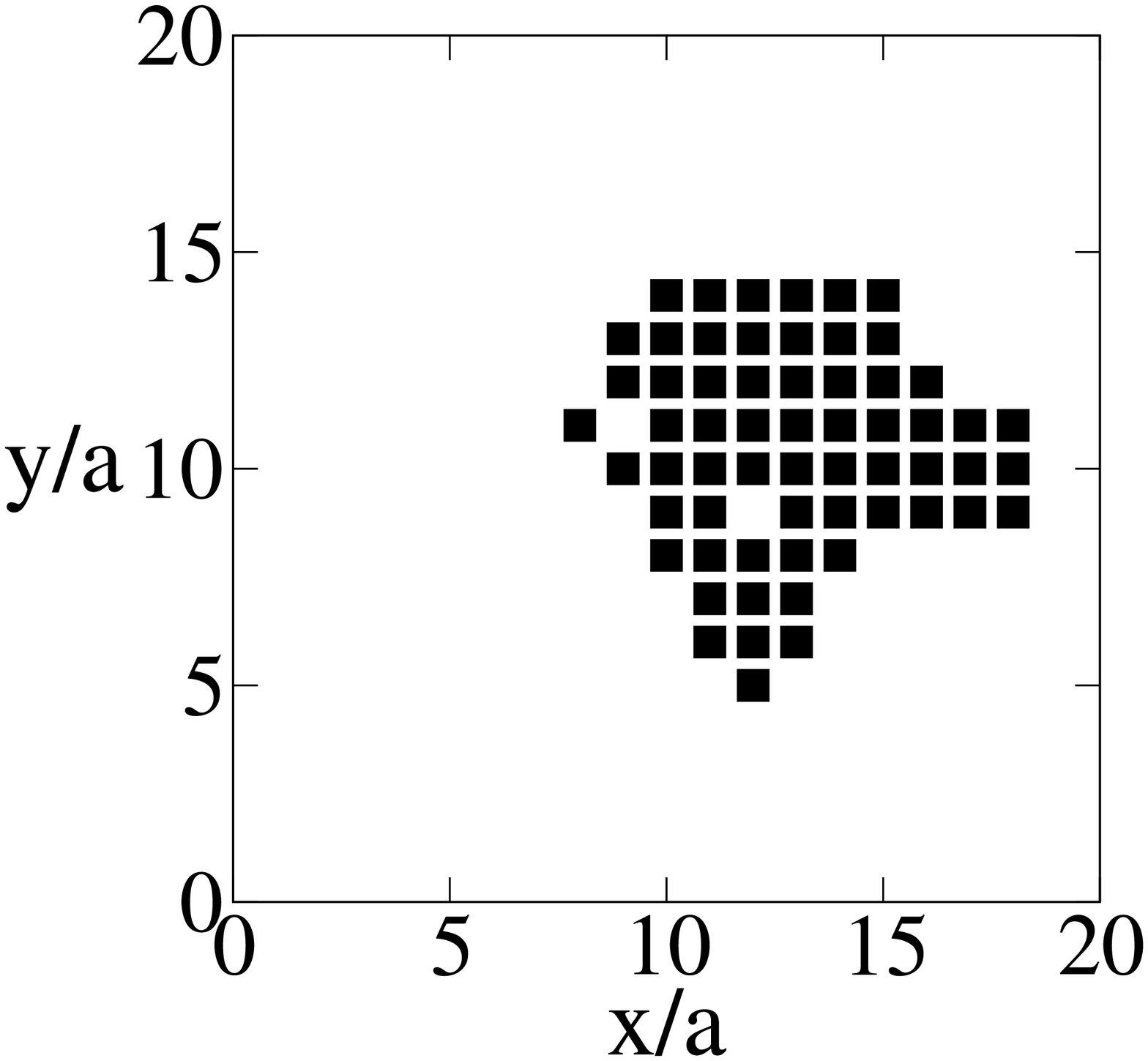}
\end{minipage}
\begin{minipage}[t]{.46 \linewidth}
 \center\epsfxsize=5.5cm \epsfbox{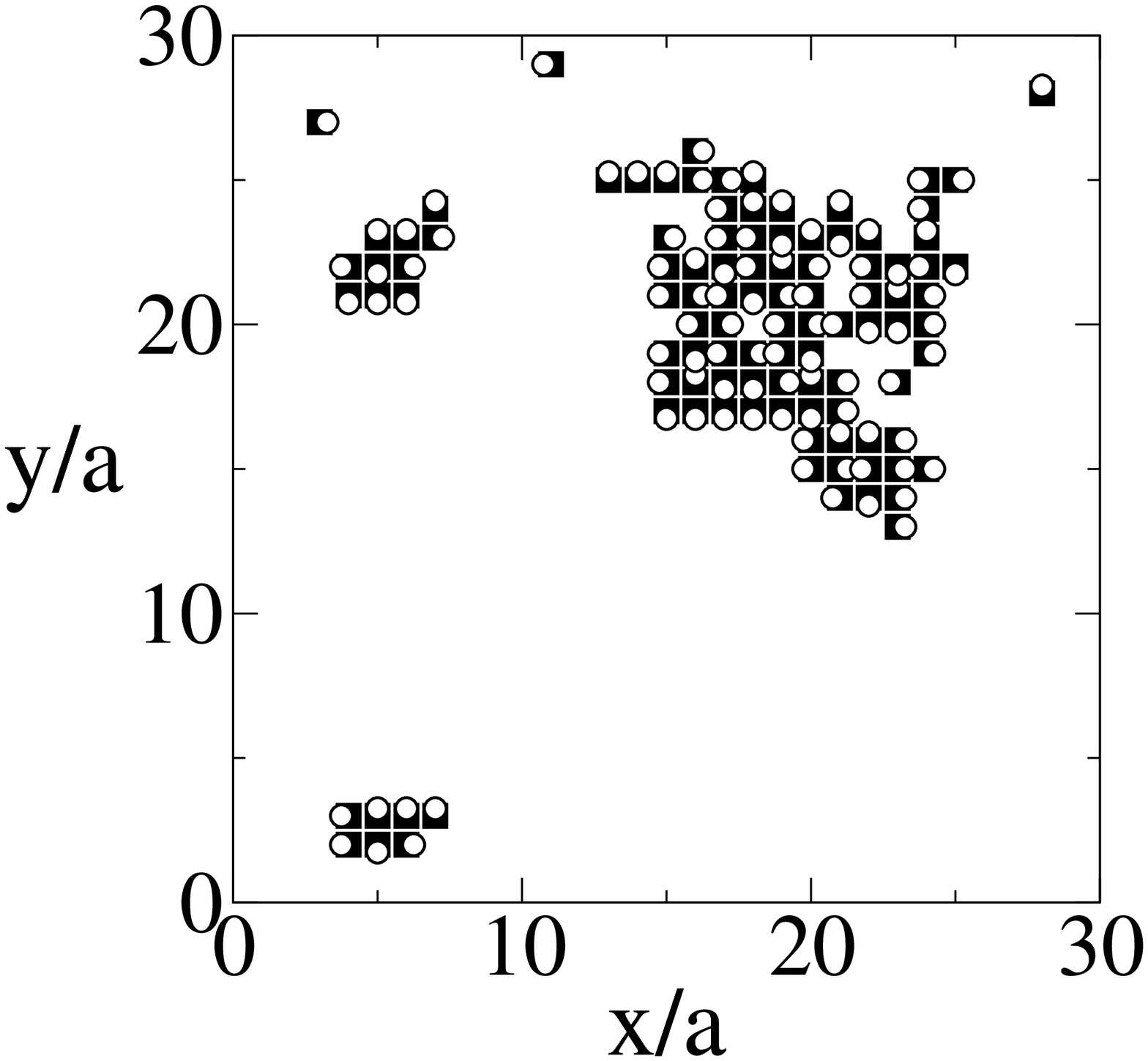}
\end{minipage}
\caption{Snapshots of 2D systems in the coexistence phase, obtained by 
Monte Carlo simulations at $T = 1.0$. Left: completely hydrophobic solute 
particles ($N_P = 0$) in water; right: mainly hydrophobic solute particles 
($N_P = 1$) in water. White circles (right) represent the polar sides of 
the solute particles, which are shown as black squares. The particles form 
compact micelles which shield the hydrophobic sides from water. The lattice 
constant $a$ is defined by the solute particle size.  }
\label{FigSnapshotMicHP0P1P}
\end{figure}

\begin{figure}[!t]
\begin{minipage}[t]{.46 \linewidth}
 \center\epsfxsize=5.5cm \epsfbox{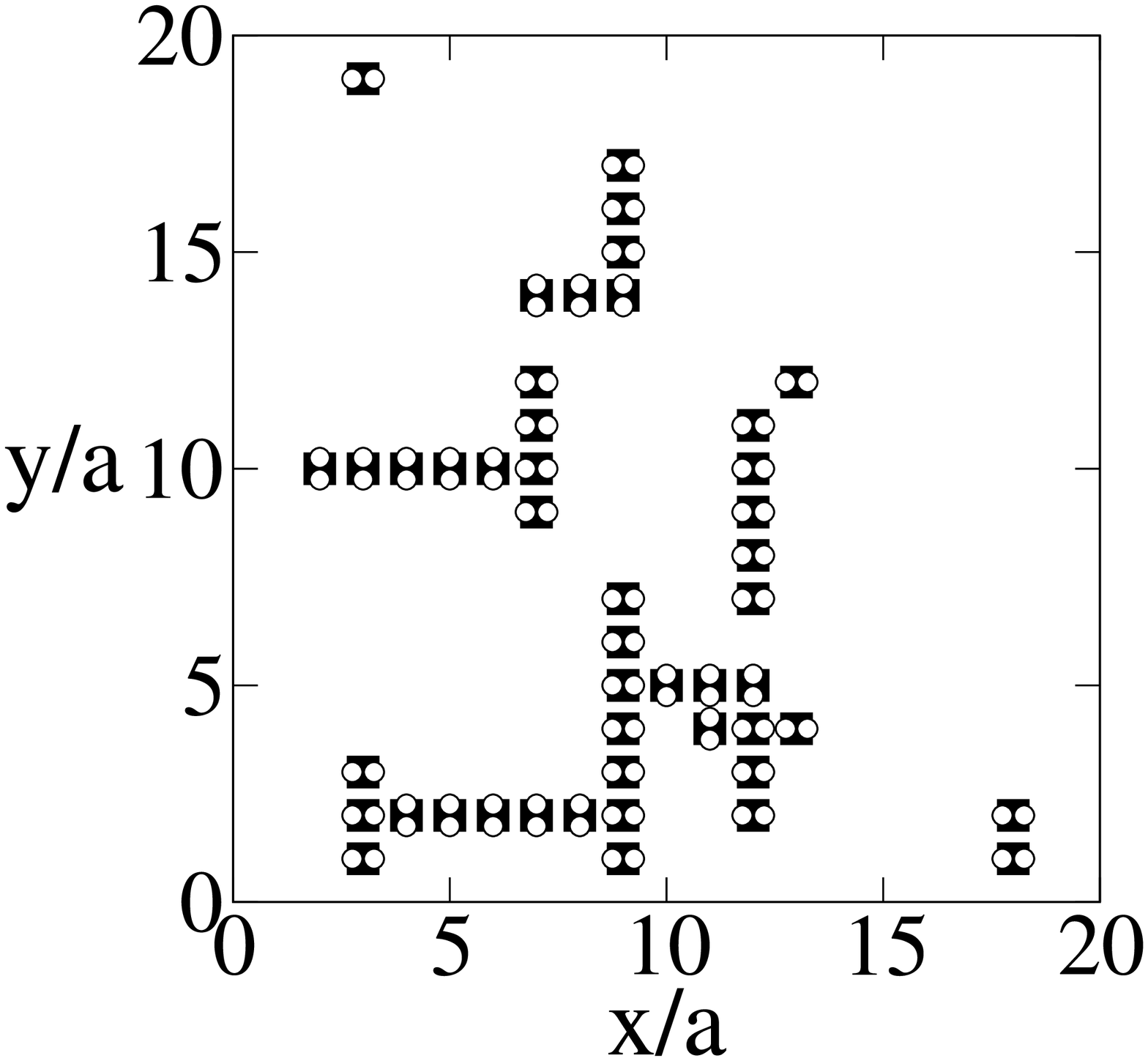}
\end{minipage}
\begin{minipage}[t]{.46 \linewidth}
 \center\epsfxsize=5.5cm \epsfbox{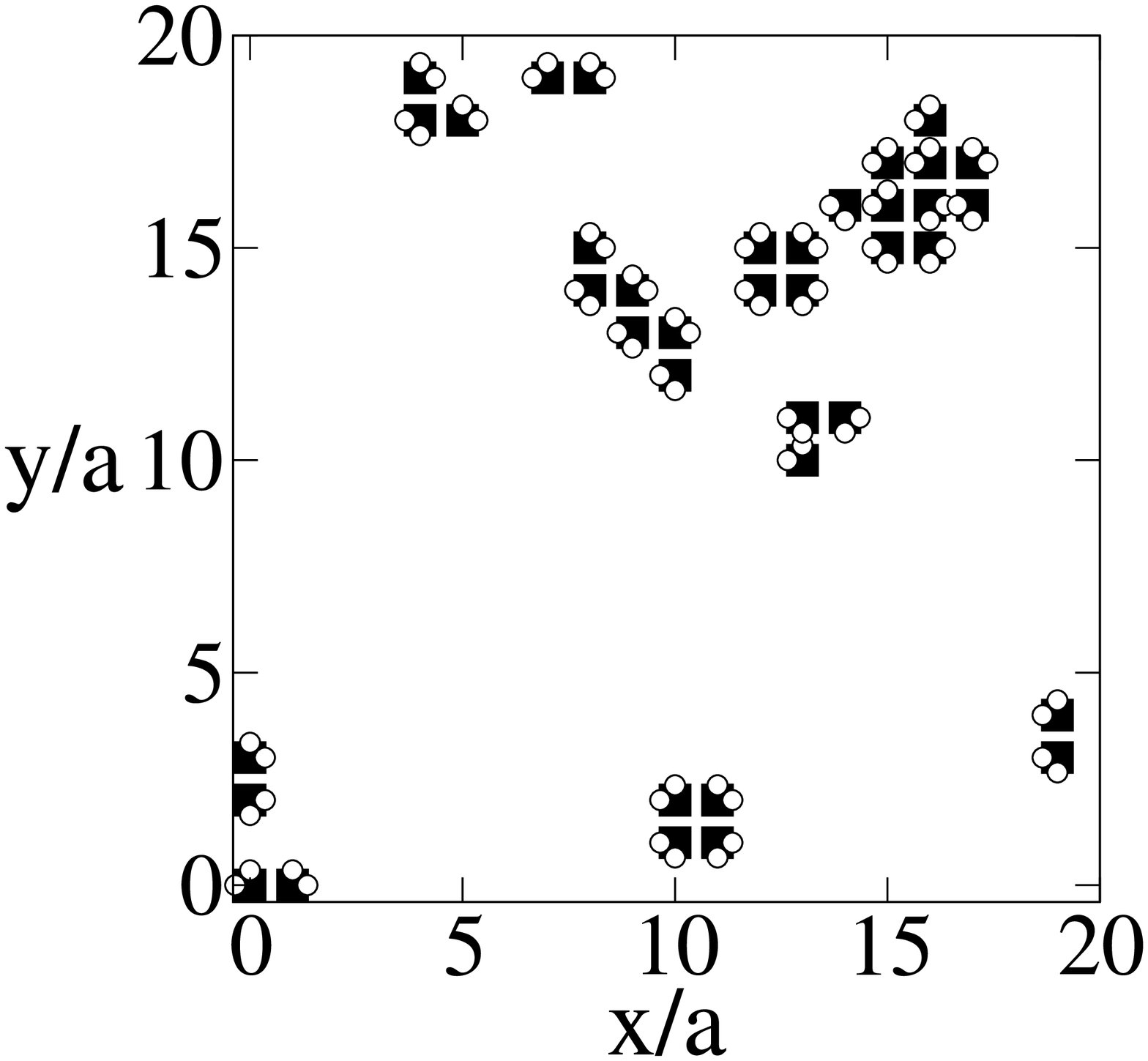}
\end{minipage}
\caption{Snapshots of 2D systems of partially hydrophobic solute particles 
($N_P = 2$) in water in the coexistence phase, obtained by Monte Carlo 
simulations at $T = 1.0$. }
\label{FigSnapshotMicHP2P}
\end{figure}

\begin{figure}[t!]
\includegraphics[width=8.5cm]{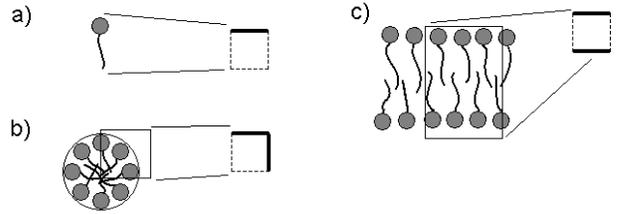}
\caption{Illustration of schematic analogs obtained using square particles 
with different arrangements of polar sides in 2D (see text). Hydrophobic 
sides of the square solute sites are shown as thick solid lines, polar 
sides as dashed lines. }
\label{FigMicPart2D}
\end{figure}

Figures \ref{FigSnapshotMicHP0P1P} and \ref{FigSnapshotMicHP2P} illustrate 
the nature of the aggregated phase using ``snapshots'' of two-dimensional 
(2D) systems at $T = 1.0$, obtained in the coexistence region for solute 
particles with different numbers of polar sides. The snapshots are taken 
after allowing the system to relax for $1\,000\,000$ steps, where every 
$20^{th}$ step is an attempt to exchange two sites and the others are 
attempts to rotate a particle. Systems containing primarily hydrophobic 
particles ($N_P = 0$ and $N_P = 1$) form mostly compact clusters, which 
minimize the number of hydrophobic surfaces exposed to the solvent. For 
lipid-like solute particles with $N_P = 1$ [Fig.~\ref{FigMicPart2D}(a)], 
the formation of perfect micelles with a hydrophobic core and a polar 
surface is prevented by the nature of the square lattice, which causes 
frustration on the edges of the micelles: an edge particle is forced to 
expose one of its hydrophobic sides to water. The model allows a lipid 
particle to occur in the core of micelles, where its polar side is in 
direct contact with the polar side of another lipid particle, because 
no distinction is made between a group of water molecules and the polar 
face of a particle. Incorporating this distinction into the description 
could be expected to generate more realistic micellar structures, albeit 
within the confines of the cubic geometry.   
 
Amphiphilic solute particles whose surface is half polar ($N_P = N_H = 2$ 
in 2D) show differing behavior depending on the polarity pattern of the 
sides. If the polar sides are adjacent on the square, small micelles 
consisting of four solute particles can be formed, which is energetically 
the most favorable configuration because no hydrophobic sides are exposed 
to water (Fig.~\ref{FigSnapshotMicHP2P}). Short, diagonal lines of 
molecules, which may be considered to represent condensed bilayers, 
can also be formed, although their ends are hydrophobic, and this 
configuration is therefore less favorable than are ``circular'' micelles 
of four solute particles. These configurations are expected from the 
construction of the sites, which is shown in Fig.~\ref{FigMicPart2D}, 
to appear as the ground states. Solute sites with two adjacent polar 
sides may be considered to represent sections of circular micelles 
[Fig.~\ref{FigMicPart2D}(b)], with the formation of small micelles as 
a consequence. In contrast, if the polar sides are opposite each other, 
the only possibility to avoid hydrophobic sides being in contact with 
water is to form lines of particles, although again the hydrophobic 
ends remain exposed to water. These squares may be taken as schematic 
representations of cross-sections of a bilayer [Fig.~\ref{FigMicPart2D}(c)].   

\begin{figure}[t!]
\includegraphics[width=8.5cm]{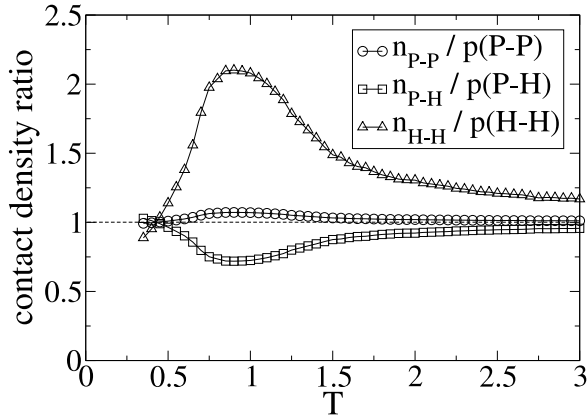}
\caption{Contact densities between polar and hydrophobic sides as a 
function of dimensionless temperature for a 3D system of solute particles 
with one polar side, $N_P = 1$. The contact densities between sides are 
normalized to the values expected in a random system for $\rho_p  = 0.24$.}
\label{FigMicRho024}
\end{figure}

The contact densities $n_{H\!-\!H}$, $n_{P\!-\!P}$, and $n_{P\!-\!H}$, 
are shown in Fig.~\ref{FigMicRho024} for a system of solute particles 
with one polar side ($N_P=1$) as a function of relative temperature. The 
results are normalized to the probability of these contacts in a randomly 
distributed system with the same particle density, $\rho_p=0.24$. At low 
temperatures, the solute particles are clearly soluble, because $n_{H\!-\!H} 
< p(H\!\!-\!\!H)$, $n_{P\!-\!P} < p(P\!\!-\!\!P)$, and $n_{P\!-\!H} > 
p(P\!\!-\!\!H)$. This reflects the formation of strongly hydrogen-bonded, 
partial cage-like structures of water molecules around the hydrophobic 
parts of the amphiphilic particles when entropy effects are minor. At 
temperatures higher than the lower critical temperature for density 
$\rho_p$, increasing entropy effects favor a screening of hydrophobic 
faces from polar ones, and the solute particles aggregate to form 
micelles. In this regime $n_{H\!-\!H} > p(H\!\!-\!\!H)$, $n_{P\!-\!P} 
> p(P\!\!-\!\!P)$, and the density of polar-hydrophobic contacts is 
suppressed, $n_{P\!-\!H} < p(P\!\!-\!\!H)$. The effect on $n_{H\!-\!H} / 
p(H\!\!-\!\!H)$ is more pronounced than $n_{P\!-\!P} / p(P\!\!-\!\!P)$ 
because the number of polar sides in the system is higher than the number 
of hydrophobic ones. Changes in $n_{H\!-\!H}$ then lead to a larger 
relative effect, and in fact the majority of $P$-$P$ contacts are intact 
even in the dissolved phase due to the high number of water molecules. 
For low particle densities, the relative effect is larger still because 
fewer $H$-$H$ contacts are possible. 

Finally, at high temperatures, the entropy becomes dominant and the 
contact densities approach their respective random values as complete 
mixing is obtained. We note that at high temperatures the contact 
densities do not recross the value $1$ to recover the low-temperature 
phase of single-particle dissolution (Fig.~\ref{FigMicRho024}). Instead 
their values simply converge to unity, implying that the complete 
miscibility takes the form of a truly random particle/water distribution.

\subsection{Lipids}
\label{secResultsLipid}

\begin{figure}[t!]
\includegraphics[width=8.5cm]{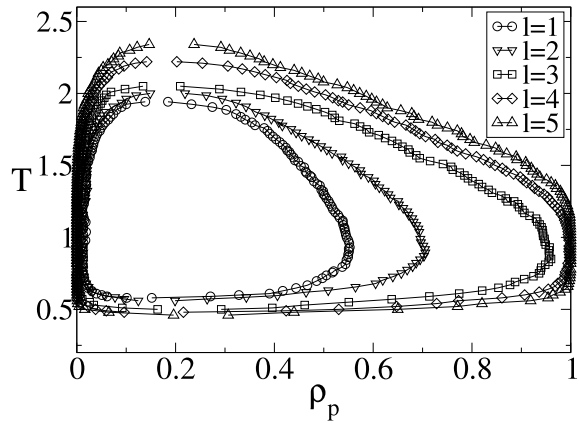}
\caption{$\rho_p$-$T$ phase diagram illustrating micelle formation in the 
extended 3D, HP model for lipid molecules of varying tail length. $l$ 
represents the relative length of the hydrophobic tail compared to the 
size of the head (Fig.~\ref{FigGlycoLipid}). The coexistence region is 
enhanced as the length of the hydrophobic tails increases, and aggregation 
is promoted. }
\label{FigBouleHPX}
\end{figure}

We have performed Monte Carlo simulations for lipid molecules in water 
on a cubic lattice for varying lengths $l$ of the hydrophobic tail, as 
described in Secs.~\ref{secLipid} and~\ref{secMethods}, to investigate 
the stability of their aggregation as a function of hydrophobicity ($l$) 
and density. The $\rho_p$-$T$ phase diagram (Fig.~\ref{FigBouleHPX}) 
shows clearly that the coexistence regime is enhanced significantly 
with increasing tail length $l$, which can be attributed to the stronger 
effective repulsive interaction between the longer hydrophobic tail of 
the lipid molecule and the surrounding water. The ratio $l$ can be taken 
as the important quantity to characterize the degree of hydrophobicity of 
a lipid molecule. Comparison with Fig.~\ref{FigGlycoLipid} indicates that 
for a typical polar head, $l = 1$ corresponds to a tail containing 
approximately four carbon atoms. Fig.~\ref{FigBouleHPX} shows that for
 lipid molecules with a tail containing approximately $8$ carbon atoms 
($l = 2$) there is already an enhancement of the coexistence region. The 
aggregation of solute particles with a tail composed of approximately $12$ 
carbon atoms ($l = 3$) is reinforced very significantly. Within our 
simplified model, at relative temperature $T = 1$ lipids with $l \geq 4$ 
are basically insoluble, forming a completely separated phase, and thus 
perfect micelles, for all densities, in rather good qualitative agreement 
with expectations based on experiments~\cite{yalkowsky}.    

\begin{figure}[t!]
\includegraphics[width=8.5cm]{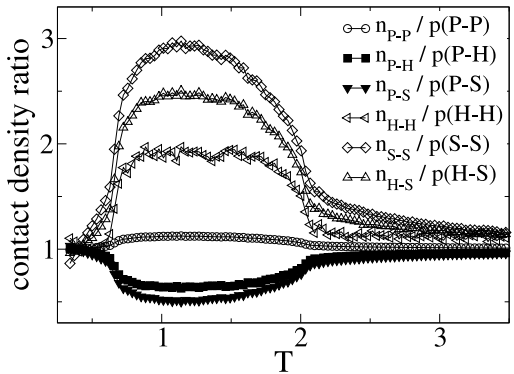}
\caption{Contact densities between the different faces as a function of 
dimensionless temperature for a 3D solution of lipid molecules with 
effective tail length $l = 3$. The contact densities between sides are 
normalized to the values expected in a random system for $\rho_p = 0.25$.   }
\label{FigHPXProba}
\end{figure}

The aggregation may be analyzed in more detail by studying the density of 
side contacts as a function of temperature. The thermal evolution of the 
density of contacts between the different sides in a system of lipid 
particles of tail length $l=3$ in water is shown in Fig.~\ref{FigHPXProba}. 
The results are normalized to the corresponding probability of these 
contacts in a randomly distributed system with the same particle density, 
$\rho_p = 0.25$. At low temperatures $n_{H\!-\!H} < p(H\!\!-\!\!H)$, 
$n_{P\!-\!P} < p(P\!\!-\!\!P)$, and $n_{S\!-\!S} < p(S\!\!-\!\!S)$, 
while $n_{P\!-\!H} > p(P\!\!-\!\!H)$, $n_{P\!-\!S} > p(P\!\!-\!\!S)$, 
and $n_{H\!-\!S} > p(H\!\!-\!\!S)$, meaning that the solute particles 
are clearly dissolved as a consequence of partial cage formation around 
the hydrophobic tails. At temperatures higher than the lower critical 
temperature for density $\rho_p$ (Fig.~\ref{FigBouleHPX}), the lipid 
particles aggregate to minimize their total exposed surface, whence 
$n_{H\!-\!H} > p(H\!\!-\!\!H)$, $n_{P\!-\!P} > p(P\!\!-\!\!P)$, and 
$n_{S\!-\!S} > p(S\!\!-\!\!S)$, while  $n_{P\!-\!H} < p(P\!\!-\!\!H)$ 
and $n_{P\!-\!S} < p(P\!\!-\!\!S)$. Because of the much stronger effect 
of an S face than of an H face on its neighboring water site, the 
normalized contact density of two S faces is highest in the aggregation 
phase and lowest below the LCST. A higher relative contact density is 
observed between S and H faces than between two H faces for the same 
reason, while below the LCST water sites prefer to form the solvation 
shell of S faces rather than of H faces. At high temperatures the contact 
densities converge, in fact rather abruptly (Fig.~\ref{FigHPXProba}) to 
their random values, indicating complete mixing.

\section{discussion}
\label{secDiscussion}
Amphiphilic molecules in aqueous solutions can form different types of 
micelles depending on their concentration and on the distribution of polar 
regions at their surfaces. Our initial investigation of the qualitative 
properties of micelle formation in a hydrophobic-polar model involved 
systematic substitution of the hydrophobic sides of cubic solute particles 
by polar ones. We determined the $\rho_p$-$T$ phase diagram for different 
surface patterns and found closed-loop coexistence curves 
(Fig.~\ref{FigBouleMicHP}), in accord with experiments using hydrophilically 
modified copolymers of poly(N-isopropylacrylamide)~\cite{rebelo}. With 
increasing polarity, the coexistence region is reduced as the solubility 
of the model amphiphiles increases. This tendency was confirmed by the same 
experiment, where the LCST of purely hydrophobic poly(N-isopropylacrylamide) 
P3 was observed to increase from $37^{\circ}$C at atmospheric pressure to 
$42$-$44^{\circ}$C when approximately $13\%$ of the monomers were 
substituted by polar species (CP2 and CP3). In our model we observe 
the same quantitative increase of $2$-$3\%$ in absolute temperature from 
$T_{LCST} = 0.545$ for purely hydrophobic solute particles to $T_{LCST} 
= 0.56$ for amphiphiles with one polar side, which represents $17\%$ of 
the particle surface. Thus the crude cubic model appears to yield good 
agreement with available data at this level of comparison. 

For particles with two polar sides the coexistence region is reduced 
dramatically. This is not surprising, considering the fact that two polar 
sides represent one third of the total particle surface, and the attractive 
interactions with water are rather strong. In amphiphilic molecules the 
polar region is usually rather small compared with the hydrophobic 
surface area. The solubility of the molecules thus increases considerably 
on substitution with polar monomers, leading to a decrease in UCST and an 
increase in LCST and CMC. In the cubic model no significant difference in 
the size of the coexistence region is observed for different distributions 
of the two sides on the solute cubes. Any further substitution of hydrophobic 
faces by polar ones results in molecules which are at least half polar: the 
solubility of such particles is always high, and thus no aggregation is 
found.

\begin{figure}[t!]
\includegraphics[width=8.5cm]{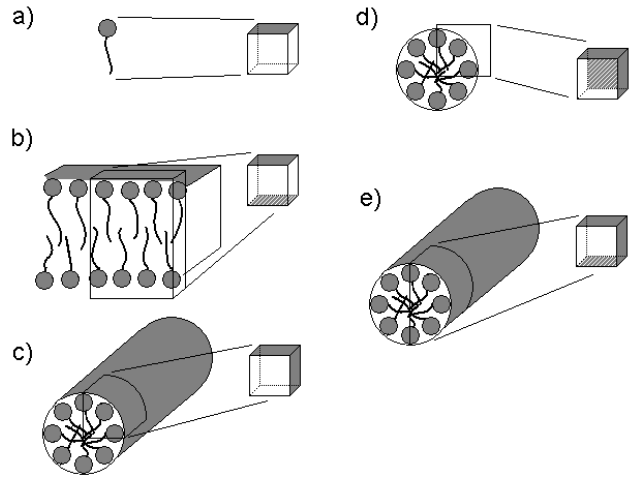}
\caption{Representation of different surface patterns on cubic solute 
particles in the HP model, and their schematic correspondence to different 
micelle types(see text). Polar surfaces and micelle segments are marked in 
gray, hydrophobic surfaces in white.}
\label{FigMicPart3D}
\end{figure}

Possible schematic interpretations of the various surface patterns of 
cubes representing solute particles in the HP model are shown in 
Fig.~\ref{FigMicPart3D}. The model is applicable for any surface pattern 
and density of solute, with the premise that one site may contain one 
amphiphilic molecule or a group of solute molecules. A given surface 
distribution of polar groups on a polymer may be characterized by a 
corresponding arrangement of polar sides on the surface of a cubic solute 
molecule. A single lipid molecule with a clear distinction between a polar 
head and a hydrophobic tail may be represented by a cube with one polar 
face [$N_P = 1$, Fig.~\ref{FigMicPart3D}(a)]. A small section of a 
cylindrical micelle would be represented by a cube with two adjacent 
polar sides [$N_P = 2$, Fig.~\ref{FigMicPart3D}(c)], while a particle 
with two opposite polar sides corresponds at the same level of approximation 
to a cross-section of a bilayer [Fig.~\ref{FigMicPart3D}(b)]. For the 
formation of ``spherical'' micelles, each site must have three adjacent 
polar sides [$N_P = 3$, Fig.~\ref{FigMicPart3D}(d)]; if two of the three 
sides are situated opposite each other [Fig.~\ref{FigMicPart3D}(e)], the 
site may again be considered as a section of a cylindrical micelle. 
Although by construction $N_P=3$ should give small spherical or 
cylindrical micelles, depending on the distribution pattern at the 
surface of each site, in fact the solubility is too high to find 
aggregates. 

Micellar structures occurring in 3D simulations are difficult to display. 
To confirm the formation of different micelle types depending on the surface 
pattern of the solute particles, we have considered snapshots of an analogous 
2D system (Figs.~\ref{FigSnapshotMicHP0P1P} and~\ref{FigSnapshotMicHP2P}). 
Here the square solute particles may be interpreted in a manner similar to 
the 3D case (Fig.~\ref{FigMicPart2D}), and the formation of small micelles 
and layers is found in the coexistence region. During the relaxation 
process, micelles grow from initial dimers to larger entities. Although 
the solute particles may rotate at a given position, they can be trapped 
in a configuration which disables the construction of perfect micelles 
or extended layers. Because there is no preference for growing a layer 
in one direction rather than in the other, short line segments are formed 
which are incompatible with others, resulting in a network of short layers. 
In the model, no distinction is made between the polar side of a water 
molecule and that of a solute particle, and a $P$-$P$ contact contributes 
the same energy independent of the molecules to which the sides under 
consideration belong. Such contacts between the polar sides of solute 
molecules are found in the interior of a micelle, which also influences 
the formation of perfect micelles. The non-zero temperature in the 
coexistence region, and the observation that the upper critical density 
is much smaller than unity, might further imply the formation of imperfect 
micelles. In fact the extent to which the upper critical density is 
significant remains unclear, because shell water sites in the model may 
be considered as belonging to the micelle phase rather than to the pure 
water phase, which would explain the low density of the organic phase 
even for perfect micelles containing no water molecules.  

We have extended our analysis to describe lipids, which represent a 
particular type of amphiphilic molecule. Lipids are distinguished by a 
special partition of the polar and hydrophobic segments along the molecule: 
a typical lipid molecule consists of a polar head and one or two hydrophobic 
tails. To incorporate these geometrical features in the model we have adapted 
the energy levels and their degeneracies according to the tail length $l$, 
where a typical lipid molecules exhibiting surface-active properties would 
be represented by values $2 \le l \le 5$~\cite{yalkowsky}. 

As expected from experiments~\cite{lang}, Monte Carlo simulations of lipids 
of increasing length illustrate a significant enhancement of the coexistence 
region (Fig.~\ref{FigBouleHPX}). Lipids with a longer tail have more 
pronounced characteristics of hydrophobic solute particles than do lipids 
where the polar head represents a considerable fraction of the molecule. 
The solubility of long-tailed lipids is therefore lower than that of 
short-tailed ones, causing a decrease in LCST, an increase in UCST, and  
a decreased CMC at any given temperature as the tail becomes longer. For 
the parameters used in the model we find that an increase from $l = 1$ to 
$l = 2$ already enhances the coexistence region substantially, and for 
that $l \geq 4$ the system is completely separated into pure water and 
essentially perfect micelles of density $\rho_{p}=1$ over a wide 
temperature range, which is fully consistent with experiment~\cite{yalkowsky}. 

A quantitative measure of micelle formation in amphiphilic and lipid 
systems is provided by the contact densities $n_{\alpha\!-\!\alpha}$ 
and $n_{\alpha\!-\!\beta}$, normalized by the corresponding probabilities 
of the contacts in a random system (Figs.~\ref{FigMicRho024}, 
\ref{FigHPXProba}). We have found fewer contacts between hydrophobic 
faces and more hydrophobic-polar contacts than would be expected for 
a random distribution below the LCST, indicating highly dissolved solute 
molecules. Above the LCST this picture is inverted, thus confirming the 
aggregation of amphiphiles and lipids, and the separation of the solution 
into two phases. From a knowledge of the system geometry and densities, 
the values of the contact ratios may also be used to confirm the extent 
of solution or aggregation, and also the effective ``purity'' of the 
micelles which form for different polar distributions on the solute 
molecules. At temperatures in excess of the UCST, the solution approaches 
a random mixture due to dominant entropy effects; in Monte Carlo studies 
the relative contact densities approach unity continuously, rather than 
undergoing a sharp transition.
 
We close by emphasizing again the limits of our analysis. We have 
formulated a model for an aqueous solution using a minimal set of 
assumptions; we have used an extremely crude representation of amphiphilic 
molecules, and have considered only a cubic system as the foundation on 
which our ``micelles'' are constrained to form. Nevertheless, we have 
obtained a realistic set of aggregation phenomena and a surprising degree 
of agreement with the available experimental results. There are, however, 
several examples of phenomena which are beyond the reach of the model at 
its current level of refinement. A solution of lipids in water may produce 
a lamellar phase of bilayers at low temperature and rather high 
densities~\cite{gompper}. Bilayer formation requires the possibility 
of smooth curvature and high flexibility~\cite{israelachvili}, and is 
precluded in the Monte Carlo simulations by geometrical constraints 
presented by the lattice. For the same geometrical reasons it is also 
difficult to find well-formed micelles in the more general case of 
amphiphiles in water. 

A further limitation is that the model considers explicitly the energy 
states of water sites, and therefore no distinction is made among P-P, H-H, 
S-H, and S-S contacts. Because there is also no difference between a group 
of water molecules and the polar face of a solute particle, the model a
llows amphiphiles to occur in the core of micelles, where their polar 
sides are in direct contact with those of other amphiphiles. For lipids, 
an S-H contact which is formed during the Monte Carlo simulation is as 
favorable as an S-S contact, although it is more likely to be broken in 
a later step. In addition, the orientation of a lipid molecule is 
irrelevant for the formation of an S-S contact, which prevents efficient 
alignment of the heads and may further hinder the formation of extended 
bilayers. The incorporation of these distinctions in a more sophisticated 
description of the lipid solution may be expected to reproduce further 
detailed properties of real systems. 

In summary, we have extended the MLG framework to include the solvation 
of amphiphilic solutes in water. Within a cubic HP model we have 
found the aggregation of solute particles, and the formation of various 
types of micelle as a function of the distribution of hydrophobic 
regions. By successive substitution of hydrophobic sides by polar ones, 
we have studied the aggregation behavior and the influence of the degree 
of hydrophobicity on the upper and lower critical solution temperatures. 
We have refined this model to describe lipid molecules of varying length, 
by adapting the interaction of the hydrophobic tail to include a 
corresponding number of neighboring water molecules, and have demonstrated 
the enhanced stability of aggregates with increasing tail length (increasing 
hydrophobicity). We have shown that primary features of micelle formation, 
which are often attributed solely to the amphiphilic nature of the solute 
particles under consideration, are reproduced by our extension of the 
solvent-based MLG model to describe alterations of water structure in 
the vicinity of the different surface regions of dissolved amphiphiles. \\
\acknowledgments

We are grateful to the Swiss National Science Foundation for financial 
support through grants FNRS 21-61397.00 and 2000-67886.02.

\end{document}